\documentclass[aps,twocolumn,superscriptaddress,showpacs]{revtex4-1}

\usepackage{graphicx}
\usepackage{dcolumn}
\usepackage{bm}
\usepackage{amsmath}
\usepackage{color}
\usepackage{ulem}
\usepackage{threeparttable}
\usepackage{enumerate}
\usepackage{paralist}

\begin{document}

\title{Thermal Transport of Fractionalized Antiferromagnetic and Field Induced States in the Kitaev Material Na$_2$Co$_2$TeO$_6$}

\author{S. K. Guang}
\thanks{These authors contributed equally to this work.}
\affiliation{Department of Physics and Key Laboratory of Strongly-Coupled Quantum Matter Physics (CAS), University of Science and Technology of China, Hefei, Anhui 230026, China}

\author{N. Li}
\thanks{These authors contributed equally to this work.}
\affiliation{Hefei National Research Center for Physical Sciences at Microscale, University of Science and Technology of China, Hefei, Anhui 230026, China}

\author{R. L. Luo}
\affiliation{Department of Physics and HKU-UCAS Joint Institute for Theoretical and Computational Physics at Hong Kong, The University of Hong Kong, Hong Kong, China}

\author{Q. Huang}
\affiliation{Department of Physics and Astronomy, University of Tennessee, Knoxville, Tennessee 37996-1200, USA}

\author{Y. Y. Wang}
\affiliation{Institute of Physical Science and Information Technology, Anhui University, Hefei, Anhui 230601, China}

\author{X. Y. Yue}
\affiliation{Institute of Physical Science and Information Technology, Anhui University, Hefei, Anhui 230601, China}

\author{K. Xia}
\affiliation{Department of Physics and Key Laboratory of Strongly-Coupled Quantum Matter Physics (CAS), University of Science and Technology of China, Hefei, Anhui 230026, China}

\author{Q. J. Li}
\affiliation{School of Physics and Optoelectronic Engineering, Anhui University, Hefei, Anhui 230601, China}

\author{X. Zhao}
\affiliation{School of Physical Sciences, University of Science and Technology of China, Hefei, Anhui 230026, China}

\author{G. Chen}
\affiliation{Department of Physics and HKU-UCAS Joint Institute for Theoretical and Computational Physics at Hong Kong, The University of Hong Kong, Hong Kong, China}

\author{H. D. Zhou}
\email{hzhou10@utk.edu}
\affiliation{Department of Physics and Astronomy, University of Tennessee, Knoxville, Tennessee 37996-1200, USA}

\author{X. F. Sun}
\email{xfsun@ustc.edu.cn}
\affiliation{Department of Physics and Key Laboratory of Strongly-Coupled Quantum Matter Physics (CAS), University of Science and Technology of China, Hefei, Anhui 230026, China}
\affiliation{Institute of Physical Science and Information Technology, Anhui University, Hefei, Anhui 230601, China}

\date{\today}

\begin{abstract}

We report an in-plane thermal transport study of the honeycomb Kitaev material Na$_2$Co$_2$TeO$_6$ at subKelvin temperatures. In zero field, the $\kappa(T)$ displays a rather weak $T$-dependence but has a non-zero residual term $\kappa_0/T$, indicating strong phonon scattering by magnetic excitation and the possibility of itinerant spinon-like excitations coexisting with an antiferromagnetic order below 27 K. We propose the zero-field ground state is a novel fractionalized antiferromagnetic (AF*) state with both magnetic order and fractionalized excitations. With both the heat current and external field along the $a*$ (Co-Co bond) direction, the $\kappa_{a*}$ exhibits two sharp minima at 7.5 T and 10 T, and its value at 8.5 T is almost the same as the pure phononic transport for the high-field polarized state. This confirms the phase boundaries of the reported field-induced intermediate state and suggest its gapless continuum excitations possibly transport heat. No such intermediate phase was found in the $\kappa_a$ for the current and field along the $a$ (zigzag chain) direction. Finally, Na$_2$Co$_2$TeO$_6$ displays a strongly anisotropic magneto-thermal conductivity since the in-plane (out-of-plane) field strongly enhances (suppresses) the $\kappa_{a*}$ and $\kappa_a$.

\end{abstract}

\maketitle

The Mott insulators with substantial spin-orbit coupling are promising candidates to realize novel and exotic quantum phases that can be difficult to achieve without the spin-orbit coupling \cite{Witczak-Krempa}. These physics are particularly relevant for many $4d$/$5d$ transition metal oxides and even for the $4f$ rare-earth compounds. Interesting quantum phases such as quantum spin ice, magnetic multipolar orders, quantum spin liquids, topological Mott insulator, {\it et al.}, have been proposed and discussed among these materials \cite{QSL1, QSL2, Gingras, Pesin, LiuChen, Hirai, LiChen}. In comparison with these heavy ions, the spin-orbit coupling of the $3d$ transition metal ions is usually much weaker and thus much less discussed. Most often, the spin-orbit coupling is responsible for the generation of weak anisotropies such as the Dzyaloshinskii-Moriya interaction and the single-ion spin anisotropies, and can be neglected in most cases \cite{Book}. In certain cases, however, the spin-orbit coupling could play an interesting and sometimes indispensable role in the $3d$ transition metal compounds. For Mott insulators, this scenario was suggested to occur for the partially filled $t_{2g}$ shell of the Ni$^{2+}$ ions in the tetrahedral crystal field environment \cite{FYLiChen} and the Co$^{2+}$ ions in the octahedral crystal field environment \cite{Liu1, Sano, Liu2}, where the spin-orbit coupling is active in the linear order, and the partially filled $e_g$ shell of the Fe$^{2+}$ ions in the tetrahedral crystal field environment \cite{GChen1, GChen2}, where the spin-orbit coupling is active in the quadratic order. Due to the spin-orbit entanglement and the effective $J =$ 1/2 local moments, the Co-based honeycomb lattice antiferromagnets, Na$_2$Co$_2$TeO$_6$, Na$_3$Co$_2$SbO$_6$, and BaCo$_2$(AsO$_4$)$_2$, were proposed as candidate Kitaev materials beyond the original $4d$/$5d$ contexts \cite{Liu1, Sano, Liu2, Zhong, Kitaev}. In this work, we investigate the thermal transport of Na$_2$Co$_2$TeO$_6$ at ultralow temperatures and provide some understanding of the low-energy excitations and the field-driven magnetic phases.

Compared with the $4d$/$5d$ compounds like $A_2$IrO$_3$ ($A =$ Na, Li) and $\alpha$-RuCl$_3$, the high-spin $3d^7$ configuration of the Co ions can induce a cancellation mechanism for the nearest-neighbor Heisenberg interactions \cite{Liu1, Sano, Liu2}. Na$_2$Co$_2$TeO$_6$ does not have the monoclinic distortion of the 4$d$/5$d$ Kitaev materials \cite{Lefrancois, Bera, Viciu}. Like $\alpha$-RuCl$_3$, Na$_2$Co$_2$TeO$_6$ also develops an antiferromagnetic (AF) order at low temperatures below $T_N \sim$ 27 K. This magnetic order was suggested to be either a zigzag \cite{Lefrancois, Bera} or three-$q$ \cite{Chen2, Lee} antiferromagnetic order. Moreover, the in-plane magnetic field can induce a successive phase transition, and an intermediate magnetic state exists between the ground state antiferromagnetic order and the spin polarized state \cite{Lin, Yao}.

Thermal transport measurements were suggested to be crucial to characterize the properties of the low-energy excitations of various quantum magnets and have been used to probe the spinon excitations in various spin liquid candidate materials. At present, only few spin liquid candidates exhibit the residual thermal conductivity at zero temperature limit, $\kappa_0/T$, a fingerprint of the itinerant fermionic excitations \cite{Yamashita, Murayama, Li2, Rao}. The previous ultralow-temperature thermal conductivity experiments on $\alpha$-RuCl$_3$ indicated that there is no spinon transport in either zero field or high magnetic field \cite{Yu}. The observations of the quantum oscillation in thermal conductivity and the half quantized thermal Hall conductivity in the proposed field-induced spin liquid of $\alpha$-RuCl$_3$, however, support the exotic spinon-like excitations \cite{Czajka, Kasahara, Yokoi}. In contrast, Na$_2$Co$_2$TeO$_6$ was reported to have the dominant phononic thermal conductivity with a significant spin-phonon scattering at the ordinary low temperatures \cite{Hong}, and the thermal Hall effect measurements at not very low temperatures revealed the magnon heat transport \cite{Li1, Yang, Takeda}. To further reveal the nature of the low-energy excitations, it is necessary to investigate the heat transport of this material at ultralow temperatures.

High-quality single crystals of Na$_2$Co$_2$TeO$_6$ were synthesized by a conventional solid reaction as previous reports \cite{Yao, Xiao1}. Polycrystalline sample was mixed with the flux of Na$_2$O and TeO$_2$ in a molar ratio of 1: 0.5: 2 and gradually heated to 900 $^{\circ}$C at 3 $^{\circ}$C/min in the air after grinding. The sample was retained at 900 $^{\circ}$C for 30 hours and was cooled to 500 $^{\circ}$C at the rate of 3 $^{\circ}$C/h. The furnace was then shut down to cool to room temperature. The as-grown single crystals are thin plates with hexagonal geometry and size up to 10 $\times$ 5 $\times$ 0.06 mm$^3$. The orientation of the crystals was confirmed by X-ray Laue back diffraction measurement. Thermal conductivity was measured by using a ``one heater, two thermometers" technique in a $^3$He/$^4$He dilution refrigerator at 70 mK -- 1.2 K, equipped with a 14 T superconducting magnet \cite{Li2, Rao}. Two rectangular shaped samples were cut from as-grown crystals. Sample A has the dimension of 4.92 $\times$ 1.56 $\times$ 0.051 mm$^3$ with the length along the $a$ axis (zigzag chain), while sample B has the dimension of 5.17 $\times$ 1.56 $\times$ 0.056 mm$^3$ with the length along the $a$* axis (Co-Co bond). The heat current was applied along the longest dimension of samples. The magnetic fields were applied along either the longest dimension of the samples (the $a$ or $a$* axis) or the shortest one (the $c$ axis). The temperature gradients were measured by two in-situ calibrated RuO$_2$ thermometers. The field dependence of $\kappa$ was measured with zero-field cooling from 30 K (above $T_N$) to the fixed temperatures.

\begin{figure}
\includegraphics[clip,width=6.5cm]{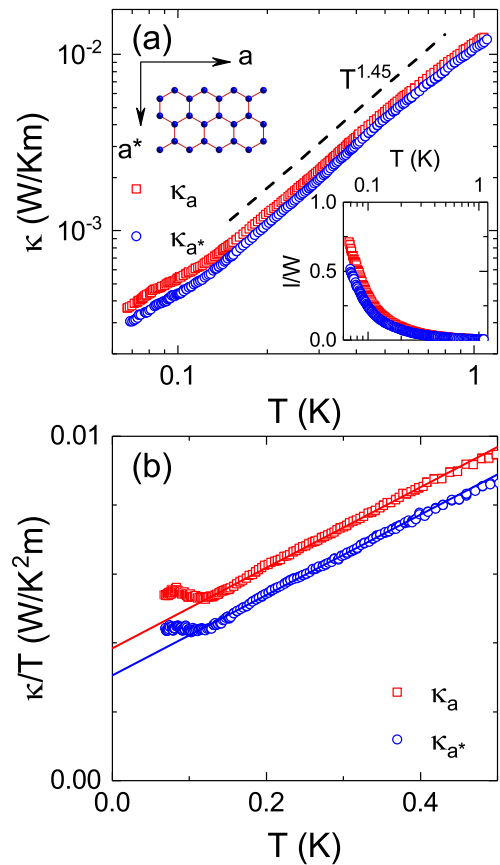}
\caption{(a) Temperature dependence of the thermal conductivity $\kappa_a$ and $\kappa_{a*}$, measured in zero magnetic field and with heat current along the $a$ axis (the zigzag direction) and the $a*$ axis (the Co-Co bond direction), respectively. The dashed line indicates a $T^{1.45}$ dependence of $\kappa$ at temperature range of 150--600 mK. The insets of (a) demonstrates the direction of $a$ axis and the $a*$ axis, and the ratio of estimated phonon mean free path to the averaged sample width. (b) Low-temperature data plotted with $\kappa/T$ vs $T$. The solid lines are linear fittings to $\kappa/T = \kappa_0/T + bT$ for $120 < T <$ 400 mK.}
\end{figure}

In Figure 1(a), we depict the temperature dependence of the thermal conductivity $\kappa_a$ and $\kappa_{a*}$, measured in zero magnetic field and with heat current along the $a$ axis (the zigzag direction, sample A) and the $a*$ axis (the Co-Co bond direction, sample B), respectively. It is known that Na$_2$Co$_2$TeO$_6$ enters a zigzag \cite{Lefrancois, Bera} or triple-$q$ \cite{Chen2, Lee} AF order below $T_N \sim$ 27 K, followed by two possible spin re-orientation transitions around 16 K and 6 K \cite{Lefrancois, Viciu, Yao, Xiao2}. In the present measurements, the temperature range is much lower than $T_N$. The magnitudes of $\kappa_a$ and $\kappa_{a*}$ are comparable, indicating nearly isotropic heat transport in the honeycomb layer. The $\kappa(T)$ shows a roughly $T^{1.45}$ behavior in a temperature of 150--600 mK, as shown in Fig. 1(a), which  strongly deviates from the standard $T^3$ behavior for either phonon or magnon thermal conductivity at very low temperatures \cite{Berman}. In addition, assuming that the thermal conductivity is purely phononic, the phonon mean free path can be calculated \cite{ZY1}. The inset to Fig. 1 shows the ratios of the phonon mean free path to the averaged sample width, which are smaller than 1 in all temperature range and indicate the existence of microscopic phonon scattering effect at such low temperatures. Apparently, there is strong scattering between the phonons and magnetic excitations in zero field even at such low temperatures.

Meanwhile, the temperature dependence of $\kappa$ at ultralow temperatures is somewhat different. Figure 1(b) shows the ultralow temperature data plotted with $\kappa/T$ vs $T$. It can be seen that the data can be linearly fitted in a rather broad temperature range of 120--400 mK, with small but non-zero residual term $\kappa_0/T$ of 0.0038 and 0.0030 W/K$^2$m for $\kappa_a$ and $\kappa_{a*}$, respectively. In quantum magnets, the thermal conductivity at ultralow temperatures can often be fitted to ${\kappa/T = \kappa_0/T + bT^{\alpha-1}}$, in which the two terms represent contributions from the itinerant fermionic magnetic excitations and phonons, respectively \cite{Yamashita, Murayama, Li2, Rao}. The power $\alpha$ can be equal to 3 under the boundary scattering limit, or smaller than 3 due to the phonon reflection at the sample surfaces or the spin-phonon scattering. We further note that the $\kappa/T$ is nearly $T$-independent at $T <$ 120 mK and points to larger residual terms, which could be due to the recovery of thermal conductivity by the itinerant excitations at extremely low temperatures. It is very surprising to observe such a non-zero residual term for Na$_2$Co$_2$TeO$_6$ since it has an AF order with gapped magnons at zero field. We interpret this result by proposing the ground state of Na$_2$Co$_2$TeO$_6$ as a fractionalized antiferromagnet (AF*), where there exist both AF order and the fractionalized spinon-like excitation in the system. Such a state was actually first proposed theoretically in the context of high-temperature superconducting cuprates, whose relevance is still unclear \cite{Lannert}. The presence of a residual thermal conductance indicates the presence of a Fermi surface or Dirac spectra of the fractionalized particles. Based on the inelastic neutron scattering result that have linearly dispersive spectral weights around the $M$ and $\Gamma$ points \cite{Chen2}, it is more natural to expect Dirac spectra for the fractionalized particles and the $M$ ($\Gamma$) point scattering corresponds to the inter-(intra-) Dirac cone scattering. The residual thermal conductance can be the universal thermal conductance associated with Dirac cones \cite{Durst}.

\begin{figure*}
\includegraphics[clip,width=17cm]{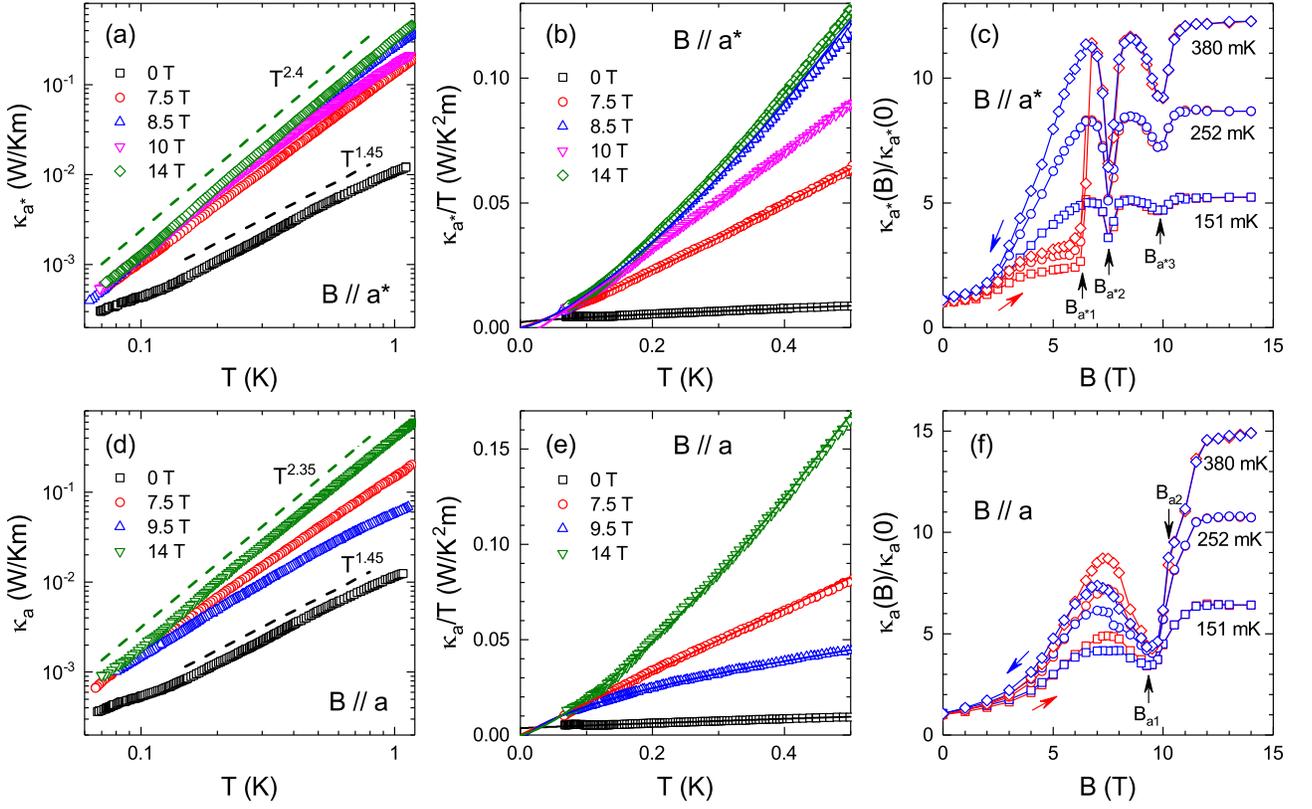}
\caption{(a) Temperature dependence of the thermal conductivity $\kappa_{a*}$ for different magnetic fields along the $a*$ axis. The dashed lines indicates the $T^{1.45}$- and $T^{2.4}$-dependence of $\kappa_{a*}$ in 0 and 14 T. (b) Low-temperature data plotted with $\kappa_{a*}/T$ vs $T$. The solid lines are fittings to $\kappa_{a*}/T = \kappa_0/T + bT^{\alpha-1}$ for $T <$ 400 mK with $\alpha =$ 2.11, 2.39, 2.06, and 2.42 for 7.5 T, 8.5 T, 10 T, and 14 T, respectively. (c) $\kappa_{a*}$ as a function of the magnetic field at 151, 252 and 380 mK. The heat current and magnetic field were applied along the $a*$ axis. The red and blue symbols indicate the measurements with increasing and decreasing field, respectively. The black arrows indicate three critical fields $B_{a*1}$, $B_{a*2}$ and $B_{a*3}$. (d) Temperature dependence of the thermal conductivity $\kappa_a$ measured in different magnetic fields along the $a$ axis. The dashed lines indicates the $T^{1.45}$- and $T^{2.35}$-dependence of $\kappa_a$ in 0 and 14 T. (e) Low-temperature data plotted with $\kappa_a/T$ vs $T$. The solid lines are fittings to $\kappa_a/T = \kappa_0/T + bT^{\alpha-1}$ for $T <$ 400 mK with $\alpha =$ 1.99, 1.55, and 2.34 for 7.5 T, 9.5 T, and 14 T, respectively. (f) The $\kappa_a$ as a function of the magnetic field at 151, 252 and 380 mK. The heat current and magnetic field were applied along the $a$ axis. The red and blue symbols indicate the data measured with increasing and decreasing field, respectively. The black arrows indicate two critical fields $B_{a1}$ and $B_{a2}$.}
\end{figure*}

Figure 2(a) shows the temperature dependence of the thermal conductivity $\kappa_{a*}$ measured in different magnetic fields along the $a*$ axis. The $\kappa_{a*}(T)$ in magnetic fields displays larger values and stronger temperature dependence; for example, at 14 T it shows a rough $T^{2.4}$ behavior, which is rather close to the $T^3$ law. This can be due to the magnetic field suppressing magnetic excitations and weakening the spin-phonon scattering. Figure 2(b) shows the same thermal conductivity data plotted with $\kappa_{a*}/T$ vs $T$. For different magnetic fields along the $a*$ axis, the $\kappa/T = \kappa_0/T + bT^{\alpha-1}$ fitting gives zero or small negative value of $\kappa_0/T$, which means no residual term $\kappa_0/T$ in the magnetic fields. Similar results were obtained for $\kappa_a$ with applying magnetic field along the $a$ axis, as shown in Figs. 2(d) and 2(e). Figure 2(c) shows the magnetic field dependence of the thermal conductivity for heat current and magnetic field along the $a*$ axis. With increasing field, the $\kappa_{a*}$ firstly increases gradually and shows a sharp increase at $B_{a*1} \sim$ 6.25 T; subsequently, the $\kappa_{a*}(B)$ exhibits two minima at $B_{a*2} \sim$ 7.5 T and $B_{a*3} \sim$ 10 T before getting saturation. The $\kappa_{a*}$ with decreasing field is much larger than those with increasing field at $B < B_{a*1}$, displaying a large and broad hysteresis.

The rather sharp minima of $\kappa_{a*}(B)$ at $B_{a*2}$ and $B_{a*3}$ clearly indicate two magnetic transitions since a minimum of the thermal conductivity most likely results from the strong scattering of phonons by magnetic fluctuations at the critical point while approaching the magnetic phase boundary \cite{Li2, Li1, Zhao, Song, Wang}. The recent inelastic neutron scattering measurements with magnetic field along the $a*$ axis revealed a field-induced intermediate magnetic state with partially polarized spins between 7.5 T and 10 T \cite{Lin2}, which correspond to the $B_{a*2}$ and $B_{a*3}$. For this state, the signature is the coexistence of magnon and the gapless continuum mode that possibly represents spinon \cite{Lin2}. Our thermal conductivity data here further validates the phase boundaries of this phase.

It is notable that the thermal conductivity value of this intermediate phase is almost as large as its value of the polarized phase. At 151 mK, for example, the $\kappa_{a*}(B)$/$\kappa_{a*}(0)$ is 5.11 and 5.22 for $B =$ 8.5 T and 14 T, respectively. Since 8.5 T is still away from the saturation field, around 11 $\sim$ 12 T, the spin-phonon scattering should be still expected and therefore the thermal conductivity value at 8.5 T is expected to be smaller than that of 14 T with purely phononic origin. There must be other contribution to enhance the thermal conductivity value at 8.5 T, which is possibly from the heat transport of the gapless continuum excitations. It is also noticed that the fitting of $\kappa(T)$ data at 8.5 T to the formula $\kappa_a/T = \kappa_0/T + bT^{\alpha-1}$ does not yield a finite residual thermal conductivity, which seems to be incompatible with the existence of itinerant magnetic excitations. However, the temperature dependence of both the phononic and magnetic thermal conductivity could be more complicated than the usual case due to the spin-phonon scattering.Therefore, the above formula may cannot precisely describe the phononic and magnetic thermal conductivity. The disappearance of hysteresis in this intermediate phase further supports its spin disordered nature.

The observed hysteresis of $\kappa_{a*}(B)$ is unusual. Base on the previous magnetization study, the $B_{a*1}$ corresponds to a first-order transition associated with the field-induced reversal of canting moments \cite{Yao, Lin}. Although the first-order magnetic transition can induce a hysteresis in the $\kappa(B)$ curve \cite{Song, Ando, Takeya}, it usually occurs in a rather narrow field region near the transition field. The present hysteresis is so broad that it extends from $B_{a*1}$ to zero field. However, this broad hysteresis at very low temperatures may be directly related to the magnetization hysteresis, which is clearly broadened with lowering temperature (from 10 to 2 K) \cite{Yao}.

Figure 2(f) shows the magnetic field dependence of the thermal conductivity for heat current and magnetic field along the $a$ axis. With increasing field, the $\kappa_a$ firstly increases and arrives a maximum at $\sim$ 7.5 T, and then decreases and reaches a minimum at $\sim$ 9.75 T, which is marked as $B_{a1}$ in the figure. This critical field is rather close to the spin polarization transition for $B \parallel a$. For field above $B_{a1}$, the $\kappa_a$ quickly increases and finally saturates in the polarized state. It is notable that there is a kink or slope change in $\kappa_a(B)$ at 10.25--10.5 T, which we marked as $B_{a2}$. This anomaly is unexpected since the spins should be already polarized. There is an obvious hysteresis between the increasing and decreasing field data for $B < B_{a1}$.

The comparison between $\kappa_{a*}(B)$ and $\kappa_a(B)$ clearly shows that the field-induced intermediate phase only exists while $B \parallel a*$. At 151 mK, the $\kappa_a(B)$/$\kappa_a(0)$ values are 4.89 at 7.5 T (the maximum position) and 6.42 at 14 T. This fact strengths our point that the gapless spin excitations in the intermediate phase indeed enhances the thermal conductivity value while $B \parallel a*$. The previous magnetization measurements did not show any hysteresis for $B \parallel a$ \cite{Yao, Lin}, so the hysteresis in $\kappa_a(B)$ is not simply due to the possible irreversible magnetization behavior. Furthermore, this hysteresis is rather odd since the relative magnitude of $\kappa$ is different at the intermediate fields and low fields ($<$ 5.5 T). It is likely to be associated with some magnetic domain structures that can scatter phonons \cite{ZY1, ZY2}. If these domains are of the AF type, they do not induce irreversible magnetization.

\begin{figure}
\includegraphics[clip,width=8.5cm]{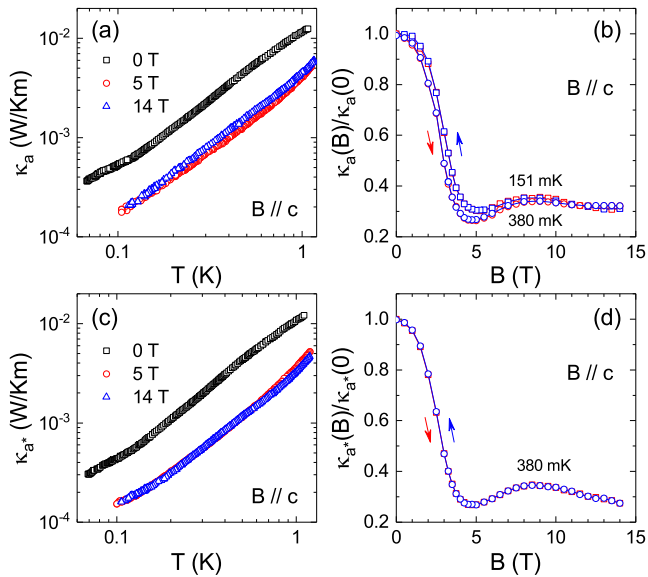}
\caption{(a) Temperature dependence of the thermal conductivity with the heat current and magnetic field along the $a$ axis and the $c$ axis, respectively. (b) Magnetic field dependence of the $\kappa_a$ at 151 and 380 mK, the red and blue symbols indicate the data for increasing and decreasing field, respectively. (c) Temperature dependence of the thermal conductivity with the heat current and magnetic field along the $a*$ axis and the $c$ axis, respectively. (d) Magnetic field dependence of the $\kappa_{a*}$ at 380 mK, the red and blue symbols indicate the data for increasing and decreasing field, respectively.}
\end{figure}

Finally, we checked the in-plane thermal conductivity measured with $B \parallel c$. Figure 3(a) shows the temperature dependence of thermal conductivity with the heat current and magnetic field along the $a$ axis and the $c$ axis, respectively. In either 5 or 14 T, $\kappa_a$ displays similar temperature dependence to that of the zero-field data while the magnitude of $\kappa_a$ is much smaller in these fields. Figure 3(b) shows the field dependence of the $\kappa_a$ for $B \parallel c$ at 151 and 380 mK. With increasing fields, the $\kappa$ decreases quickly to reach a minimum around 5 T and displays a weak field dependence up to 14 T, with a small and broad peak at $\sim$ 8 T. In addition, $\kappa_a(B)$ curves for $B \parallel c$ have no hysteresis between the increasing and decreasing field data. Figures 3(c) and 3(d) show the data for the same measurements with the heat current and magnetic field along the $a*$ axis and the $c$ axis, respectively. Both the temperature dependence and field dependence are nearly the same between the $\kappa_a$ and $\kappa_{a*}$. That is, the magnetic field along the $c$ axis induces isotropic effect on the $a$-axis and $a*$-axis heat transport properties. The field dependence of $\kappa$ for $B \parallel c$ can be due to that the magnon-phonon scattering becomes more significant in the $c$-axis fields. Whereas, the minimum at $\kappa(B)$ is likely due to two competitive contributions of magnons induced by magnetic fields. With increasing field, the magnon gap will shrink and the magnon excitations will be populated; thus, the magnon transport can enhance the $\kappa$ while the stronger magnon-phonon scattering can suppress the $\kappa$. Since the magnetization data did not show any signature of spin structure transitions at low fields for $B \parallel c$ \cite{Yao, Lin}, the valley of $\kappa(B)$ at low fields is likely due to the competition of these two effects. It should be pointed out that a recent heat transport study revealed that there is a clear hysteresis in $\kappa(B)$ for $B \parallel c$ at temperatures down to 8 K \cite{Yang}, which was ascribed to the small ferrimagnetic moment along the $c$ axis \cite{Yao}. However, the present ultralow-temperature data does not exhibit hysteresis in $\kappa(B)$. In addition, our previous $\kappa(B)$ data at 0.36 -- 1.95 K consistently did not show hysteresis \cite{Li1}. Therefore, for $B \parallel c$ the hysteresis of $\kappa(B)$ appears at high temperatures.

It is obvious that the in-plane thermal conductivity of Na$_2$Co$_2$TeO$_6$ reacts very differently to the in-plane or out-of-plane field. In general, it is enhanced by the in-plane field but suppressed by the out-of-plane field. Such anisotropic magneto-thermal conductivity is rarely observed in magnetic materials, in which the magnetic fields in different directions usually affect the thermal conductivity in the same way. To our knowledge, only NiCl$_2$-4SC(NH$_2$)$_2$ (DTN) was found to exhibit anisotropic magneto-thermal conductivity at very low temperatures \cite{Sun}. DTN is a spin-1 chain system which has a spin gapped ground state. While applied magnetic fields along the spin-chain direction closes the spin gap, DTN can enter a gapless magnon Bose-Einstein condensation (BEC) state. In this state, the magnetic field along the spin chains can enhance the thermal conductivity along the chain-direction by introducing the magnon transport; in contrast, the magnetic field perpendicular to the spin chains does not induce the magnon BEC state and suppress the thermal conductivity. However, the present Na$_2$Co$_2$TeO$_6$ result must have a different mechanism, and we think this is due to the combined effects of the anisotropic magnetic interaction and the specific magnetic structures. Moreover, the $\kappa_{a*}(B)$ and $\kappa_a(B)$ are obviously different too. First, the $\kappa_{a*}(B)$ shows the field-induced intermediate magnetic state but $\kappa_a(B)$ does not. Second, within the hysteresis, the $\kappa_{a*}(B)$ is larger with the decreasing field for $B \parallel a*$ while the $\kappa_a(B)$ is larger with the increasing field for $B \parallel a$. All these differences indicate that in the $ab$ plane, the exchange interactions along the $a*$ and $a$ directions are different from each other, which should be accounted for future studies.

In summary, our ultralow-temperature thermal conductivity data of Na$_2$Co$_2$TeO$_6$ suggests the existence of itinerant fermionic magnetic excitations in the zero magnetic field that is interpreted from the fractionalized antiferromagnet. We further show the existence of the intermediate magnetic state and suggest its gapless continuum excitations possibly transport heat. Moreover, our data show that Na$_2$Co$_2$TeO$_6$ is a rare magnet exhibiting strong anisotropic magneto-thermal conductivity. Such complex thermal conductivity behaviors reflect the unique exchange interactions in Na$_2$Co$_2$TeO$_6$, which call for further studies to be clarified.

\begin{acknowledgements}

We acknowledge very useful discussion with Yuan Li. This work was supported by the National Natural Science Foundation of China (Grant Nos. U1832209, 11874336, 12274388, 12174361, 12104011, and 12104010) and the Nature Science Foundation of Anhui Province (Grant Nos. 1908085MA09, 2108085QA22) and Hong Kong CRF. The work at the University of Tennessee was supported by NSF with Grant No. NSF-DMR-2003117.

\end{acknowledgements}

\end{document}